\documentclass[apj,numberedappendix,appendixfloats,tighten,a4paper,twocolumn]{emulateapj}
\usepackage{graphicx,amsmath,multirow,rotating}
\usepackage{hyperref}

\newcommand{\vrun}{V_{\text{run}}}
\newcommand{\rrun}{\rho_{\text{run}}}
\newcommand{\vdet}{\rho_{\text{det}}}
\newcommand{\prad}{p_{\text{rad}}}
\newcommand{\pgas}{p_{\text{gas}}}
\newcommand{\psh}{p_{\text{sh}}}
\newcommand{\rhosh}{\rho_{\text{sh}}}
\newcommand{\etsh}{\eta_{\text{sh}}}

\newcommand{\teff}{T_{\text{eff}}}
\newcommand{\rph}{r_{\text{ph}}}

\newcommand{\tdiff}{t_{\text{diff}}}
\newcommand{\tdrop}{t_{\rm drop}}

\newcommand{\Ttkev}{T_{10\rm keV}}

\newcommand{\tes}{\tau_{\rm es}}
\newcommand{\dm}{\delta M}

\newcommand{\rocrit}{\rho_{\text{crit}}}
\newcommand{\tdet}{t_{\text{det}}}

\newcommand{\DDir}{\relax{D\kern-.7em{/}}}

\newcommand{\be}{\begin{equation}}
\newcommand{\ee}{\end{equation}}
\newcommand{\bea}{\begin{equation*}}
\newcommand{\eea}{\end{equation*}}

\newcommand{\nin}{\relax{\in\kern-.8em{/}}}

\newcommand{\cm}{\mbox{ cm}}

\newcommand{\erg}{\mbox{ erg}}

\newcommand{\eV}{\mbox{ eV}}

\newcommand{\OpUnit}{\mbox{ cm}^2 \mbox{ g}^{-1} }
\newcommand{\DenUnit}{\mbox{ g}\mbox{ cm}^{-3}}
\newcommand{\VelUnit}{\mbox{ cm}\mbox{ s}^{-1}}

\newcommand{\PresUnit}{\mbox{ erg}\mbox{ cm}^{-3}}
\newcommand{\LumUnit}{\mbox{ erg}\mbox{ s}^{-1}}
\newcommand{\elem}[2]{${}^{#2}${#1}}

\newcommand{\sref}{\S~\ref}

\begin{document}
\title{Early emission from type I\lowercase{a} supernovae}
\author{Itay Rabinak\altaffilmark{1}, Eli livne\altaffilmark{2} and Eli Waxman\altaffilmark{1}}
\altaffiltext{1}{Department of Particle Physics \& Astrophysics, Weizmann Institute of Science, Rehovot 76100, Israel}
\altaffiltext{2}{Racah Institute of Physics, Hebrew University, Jerusalem, Israel}

\email{itay.rabinak@weizmann.ac.il}
\date{\today}

\begin{abstract}
A unique feature of deflagration-to-detonation (DDT) white dwarf explosion models of SNe of type Ia is the presence of a strong shock wave propagating through the outer envelope. We consider the early emission expected in such models, which is produced by the expanding shock-heated outer part of the ejecta and precedes the emission driven by radioactive decay. We expand on earlier analyses by considering the modification of the pre-detonation density profile by the weak-shocks generated during the deflagration phase, the time evolution of the opacity, and the deviation of the post-shock equation of state from that obtained for radiation pressure domination. A simple analytic model is presented and shown to provide an acceptable approximation to the results of 1D numerical DDT simulations. Our analysis predicts a $\sim10^3$~s long UV/optical flash with a luminosity of $\sim1$ to $\sim3\times 10^{39} \LumUnit$. Lower luminosity corresponds to faster (turbulent) deflagration velocity. The predicted luminosity of the UV flash is an order of magnitude lower than that of earlier estimates, and is expected to be strongly suppressed at $t>\tdrop\sim1$~hr due to the deviation from pure radiation domination.

\end{abstract}

\maketitle

\section{Introduction}

Type Ia supernova (SN Ia) explosions are commonly accepted to be driven by unstable thermonuclear burning of C/O white dwarfs. A key open question is whether the nuclear burning front propagates as a deflagration or as a detonation wave. Both deflagration models \citep{Ivanova74,Nomoto76,Nomoto84,woosley1986pse} and delayed-detonation models, in which a deflagration wave spontaneously transforms into a detonation wave 
(\citealp{Woosley90baa}\footnote{Similar to a model previously proposed by A.~M.~Khokhlov} \citealp{khokhlov1991dtm,Yamaoka92,khokhlov1993lct,Arnett94,Arnett94b,woosley1994mss,hoflich1995ddm,hoeflich1996emt,iwamoto1999ncm}),
are claimed to best fit the observational data.

One consequence of a deflagration to detonation transition (DDT) is the presence of a shock wave that erupts through the progenitor's surface. The detonation wave transforms into an ordinary shock wave, behind which thermonuclear burning does not release much energy, near the edge of the star, at densities $\sim10^6{\rm g/cm^3}$ \citep[e.g.][]{Piro10}, where the thickness of the burning layer becomes comparable to the scale height. Unique signatures of the presence of a strong shock are an X-ray outburst, which is expected to be produced when the shock breaks out from the stellar edge \citep[][]{Colgate74b,Falk78,Klein78}, and early UV-optical emission, which is produced by the expanding and cooling shock heated outer shells of the progenitor \citep[][]{Falk78} and precedes the emission powered by radioactive decay.

The observed properties of the shock breakout were investigated using analytic order of magnitude estimates \citep[e.g.][]{MM,Katz10,Piro10,Nakar10} and numerical calculations for particular progenitors \citep[e.g.][]{Ensman92,Blinnikov00,Utrobin07,Hoflich09,Tominaga09,Tolstov10,Dessart11,Kasen11}. An exact description of the time dependent radiation emission for non-relativistic shock breakout from a general progenitor (without an optically thick wind) has been recently provided in \citep{SapirKatz11,KatzSapir11}. This description is accurate during the "planar phase" of the expansion, that is as long as the distance traveled by the expanding shells is small compared to the progenitor's radius. Approximate analytic descriptions of the emission during the later "spherical phase", when the shells expand to radii much larger than the original radius of the progenitor, were given by several authors \citep[e.g.][]{Chev,Waxman07,Chev08,Rabinak10,Nakar10}. The recent analysis of \citet{Rabinak10} provides an approximate analytic description including the effects of opacity variations due to recombination, which are of particular importance for compact progenitors with hydrogen poor envelopes, and a method for inferring the relative extinction directly from the UV/O light curves.

The early emission from a WD undergoing a DDT explosion was recently discussed by \citet{Nakar10}, who gave general order of magnitude estimates, and by \citet{Piro10}, who used a more realistic description of the propagation of the detonation wave.
In this paper we improve on the existing analyses by taking into consideration three effects that have not been addressed earlier: the modification of the pre-detonation density profile by the weak-shocks generated during the deflagration phase, the time evolution of the opacity due to recombination, and the deviation of the post-shock equation of state (EOS) from that obtained assuming the pressure to be purely contributed by radiation (neglecting the plasma contribution).

A note is in place here regarding the third effect. During the "spherical phase" the radius $r$ and density $\rho$ of the radiation emitting shells at some given time $t$ is essentially determined by the expansion velocity $v$ and by the opacity $\kappa$ ($r\sim vt$, $\tau\sim\kappa\rho r$), and largely independent of the initial radius of the progenitor $R_*$. The initial post shock pressure of the radiation emitting shells is $\sim\rho_0 v^2$, where $\rho_0\sim(r/R_*)^{3}\rho$ is the initial density. This implies that the initial ratio of radiation pressure, $aT^4\sim\rho_0 v^2$, to plasma pressure, $\sim\rho_0 T/\mu$, is $\propto \rho_0^{-1/4}\propto R_*^{3/4}$. Thus, while neglecting the plasma pressure is an excellent approximation for large (core-collapse SN) progenitors, it is not a good approximation for SN Ia progenitors. As we show below, taking into account the deviation from pure radiation domination leads to strong suppression of the flux at $t>\tdrop\sim1$~hr.

The properties of the emission during both the planar phase and the spherical phase are not very sensitive to the initial density profile \citep{SapirKatz11,KatzSapir11,Rabinak10}. We therefore discuss first, in \S~\ref{sec:spherical}, the opacity and EOS effects using a simple analytic model for the spherical phase emission, assuming a power-law dependence of the progenitor's density on distance from the stellar surface and a simple self-similar description of the shock propagation (we include in \S~\ref{sec:est_t_drop} also a derivation of $\tdrop$ using a more realistic description). The planar phase, which is not much affected by the opacity and EOS effects and lasts for $\lesssim0.1$~s for the compact SN Ia progenitors, is briefly discussed in \S~\ref{sec:breakout_flash_ddt}. We then discuss, in \S~\ref{sec:sim_desc} and \S~\ref{sec:res_early_n_BO}, the effect of deviations from the simple description of \S~\ref{sec:spherical}, using detailed hydrodynamical models of DDT explosions. The numerical models are described in \sref{sec:sim_desc}, and the emission properties are described in \sref{sec:res_early_n_BO}.
Our results are summarized and discussed in \sref{sec:sum_and_disc}.

\section{Simple analytic estimates}\label{sec:early_emis_sh_heated}

\subsection{The spherical phase}
\label{sec:spherical}

For the analysis of this section we assume that the shock accelerates down the density gradient following the self-similar solution of \citet{GandelMan56,Sakurai60}, $v\propto\rho_0^{-\beta}$ with $\beta=0.19$. This description of the shock velocity is valid for explosions in which the energy is released at small radii, $r\ll R_*$ \citep[see][]{MM}, i.e. ignoring the fact that part of the energy is released by nuclear burning at large radii. The distributed energy release is accurately described in our numerical analysis (\S~\ref{sec:sim_desc} and \S~\ref{sec:res_early_n_BO}). As we show there, the simple model described in this section provides an acceptable description of the velocity profile of the ejecta, and a rough description of the pressure profile in regions where radiation dominates the pressure (see fig.~\ref{fig:FinPrsFitAsTau}).

The early, $t<1$~day, emission from SNe is dominated by the outer $\lesssim10^{-3}M_\odot$ shells of the progenitor, which are heated by the SN shock and are emitting radiation while expanding and cooling. We discuss the opacity and EOS effects in \S~\ref{sec:opct_eff_emis} and \S~\ref{sec:est_t_drop} respectively, using the analytic model of \citet{Rabinak10}. This simple model's assumptions are that the initial density profile of the progenitor is given by a power-law of the form $\rho_0\propto \delta^{n}$ where $\delta = (R_*/r_i -1)$ and $r_i$ is the initial radius of the shell, that the post-shock energy density is dominated by radiation, and that the post-shock expansion is adiabatic. Since the results are not sensitive to the value of $n$, we adopt below $n=3$ (the value obtained for non degenerate radiative envelopes).

The model assumptions are similar to those used in other analytic studies of the early emission \citep[e.g.][]{Chev,Waxman07,Chev08,Rabinak10,Nakar10}. We use the \citet{Rabinak10} analysis since it includes a realistic description of the opacity (beyond the Thomson opacity of fully ionized plasma used in other analyses). Another comment is in place here regarding the various analyses. For the calculation of the luminosity, the diffusion of radiation below the photosphere is ignored in \citet{Rabinak10} while it is taken into account in \citet{Chev08,Nakar10}. As pointed out in \citet{Rabinak10}, the effects of the diffusion on the luminosity are indeed negligible, and the results obtained with and without diffusion are nearly identical \citep[e.g. compare][]{Rabinak10,Nakar10}. The effects of diffusion are, on the other hand, important for determining the spectrum, or color temperature. The color temperatures obtained in \citet{Rabinak10} differ from those obtained in \citet{Nakar10} due to the more accurate description of the opacity.

\subsubsection{Opacity} \label{sec:opct_eff_emis}

As long as photospheric temperature is well above the recombination temperature, the ejecta is nearly fully ionized and the opacity is time independent and dominated by electron scattering. At this stage the bolometric luminosity and the effective temperature are given by (see Eqs.~(13) and (15) in \citealt{Rabinak10})
\begin{equation}\label{eq:L_n3}
    L(t) = 3.2 \times 10^{39} \frac{E^{0.85}_{51} R_{8.5} }
    {f^{0.16}_{\rho} (M/1.4 M_{\odot})^{0.69}\kappa^{0.85}_{0.2} }t_2^{-0.31} \erg \, {\rm s}^{-1},
\end{equation}
and
\begin{equation}\label{eq:T_n3}
    \teff(t) =  3.5 \, f_{\rho}^{-0.022}
    \frac{E_{51}^{0.016} R_{8.5}^{1/4} }
    {(M/1.4M_{\odot})^{0.03} \kappa^{0.27}_{0.2}}
    t_2^{-0.47} \eV.
\end{equation}
Here $t =10^2  t_2$~s is the time after breakout, $R_* = 10^{8.5} R_{8.5} \cm$, $\kappa = 0.2  \kappa_{0.2}\OpUnit$, $M$ is the ejecta mass, $E = 10^{51} E_{51}$~erg is the explosion energy and $f_{\rho}$ is a dimensionless factor that depends on the envelope density profile \citep[see][]{Rabinak10}. $\kappa_{0.2} \approx 1+X$, where $X$ is the H mass fraction. The color temperature $T_{\text{col}}$ is set by photons that can diffuse and reach the photosphere after thermalizing in an inner layer.
As a result, $T_{\text{col}}$ is higher than the effective temperature and is approximately given by $T_{\text{col}}\approx 1.2 \teff$ \citep{Rabinak10}.

When the photospheric temperature approaches the recombination temperature of the ejecta, the approximation of constant opacity is no longer valid and the decline of the opacity, which is composition dependent, has to be taken into consideration. For H poor envelopes, the decline in the opacity becomes important for $\teff \lesssim 3\eV$. The bolometric luminosity for He and C/O envelopes is given in this regime by \cite[see Eqs.~(25) and (29) in][]{Rabinak10}
\begin{equation}\label{eq:L_n3He}
    L^{[He]}(t) = 3.1\times 10^{39}
    \frac{E^{0.84}_{51} R_{8.5}^{0.85} }
    {f^{0.15}_{\rho} (M/1.4 M_{\odot})^{0.67} }t^{-0.02}_3 \erg\, {\rm s}^{-1},
\end{equation}
and
\begin{equation}\label{eq:L_n3CO}
    L^{[C/O]}(t) = 4.4\times 10^{39} \frac{E^{0.83}_{51} R^{0.8}_{8.5} }
    {f^{0.14}_{\rho} (M/1.4 M_{\odot})^{0.67}}t_3^{0.07} \erg \, {\rm s}^{-1},
\end{equation}
respectively (note, that these equation correct for a typo that appeared in Eqs.~(25) and (29) in \citet{Rabinak10}, where the subscript of $R$ should be "12" and not "13"). Here, $t=10^3t_3$~s.
The effective temperature is given by \cite[see Eqs.~(23) and (27) in][]{Rabinak10}
\begin{equation}\label{eq:T_n3He}
    \teff^{[He]}(t) = 1.5 \eV f_{\rho}^{-0.02} R_{8.5}^{0.2} t_3^{-0.38},
\end{equation}
and
\begin{equation}\label{eq:T_n3CO}
    \teff^{[C/O]}(t) = 1.6 \eV f_{\rho}^{-0.017} R_{8.5}^{0.19} t_3^{-0.35},
\end{equation}
both with very weak dependence on $E$ and $M$.

The composition of the WD's outer mass shells at the onset of the explosion is uncertain. The estimated maximum mass of H and He shells are $\lesssim 10^{-5} M_{\odot}$ \citep{Shen09a} and $\lesssim 10^{-3} M_{\odot}$ \citep{Iben89,Shen09b} respectively.
As the diffusion sphere propagates inwards to larger mass shells it may pass through shells with different compositions.
For $t_2 \lesssim 1$, the plasma in the photosphere is fully ionized for all types of compositions, and the luminosity and temperature are given by eqs.~\eqref{eq:L_n3} and \eqref{eq:T_n3} respectively. For $t_2>3$ the photosphere propagates beyond the outer $\sim 10^{-5} M_{\odot}$ and the effective temperature drops below 3~eV. At $t_2>3$ we therefore expect the radiation to be emitted from shells dominated by He or C/O, with luminosity and temperature given by eqs.~\eqref{eq:L_n3He} and~\eqref{eq:T_n3He} or \eqref{eq:L_n3CO} and \eqref{eq:T_n3CO}, depending on the composition. For $ 1 \lesssim t_2 \lesssim 3$, the detailed properties of the emission depend on the amount of H ($X$ as function of mass). A comparison of eqs.~\eqref{eq:L_n3} and~\eqref{eq:L_n3He} or \eqref{eq:L_n3CO} implies that the modification of the opacity with time leads to a nearly time independent luminosity at $t$ larger than few hundred seconds.

\subsubsection{EOS}\label{sec:est_t_drop}

As long as the plasma pressure, $p$, is dominated by radiation, and the evolution is adiabatic, the radiation pressure of a given fluid element drops like $p=\prad\propto\rho^{4/3}\propto r^{-4}$. For shells at which the thermal energy density is dominated by the plasma thermal energy we have $p=\pgas\propto\rho^{5/3}\propto r^{-5}$, which implies, for radiation in thermal equilibrium, $\prad\propto\rho^{8/3}\propto r^{-8}$. At the late stages in which we are interested the radiation may not be in thermal equilibrium, but is expected to be in Compton equilibrium (the number of collisions required to modify the photon energy is $\sim c/v$ while the number of collisions a photon undergoes over a dynamical time is $\sim \tau c/v$). For Compton equilibrium, the radiation energy density drops like $\prad\propto\pgas\propto\rho^{5/3}\propto r^{-5}$. Thus, at times at which radiation is emitted from shells at which the thermal energy density is no longer dominated by radiation, the radiation energy density in the emitting shells and the luminosity are expected to be suppressed by a factor $\sim (\rho/ \rho_0)^{1/3}\sim(r/R_*)^{-1}$ to $\sim (\rho/ \rho_0)^{4/3}\sim(r/R_*)^{-4}$ compared to the luminosity expected assuming radiation domination (i.e. compared to the luminosity given in \S~\ref{sec:opct_eff_emis}). In what follows we estimate the time $\tdrop$, defined as the time at which the diffusion front reaches layers in which the ideal gas pressure is equal to the radiation pressure. For $t>\tdrop$ a strong suppression of the luminosity is expected.

Assuming that the post shock pressure is dominated by radiation ($\prad = a T^4 / 3$, where $T$ is the temperature and $a$ is the radiation constant) and that the plasma pressure is given by $\pgas = \rhosh T / \mu m_p $ (where $m_p$ is the proton mass and  $\mu$ is the molecular weight), and using the self-similar shock description \citep[assuming an adiabatic index of $\gamma= 4/3$, e.g.][for details]{Rabinak10}, the post-shock ratio of radiation to gas pressure is given by
\begin{equation}\label{eq:def_eta_Rab10}
     \etsh \equiv
     \frac{\prad}{\pgas}  =
     \frac{1.2\times 10^3 E_{51}^{3/4} (\mu/ 2)}
     {(\rho_0/\DenUnit)^{0.53} R^{0.84}_{8.5}
     (M/1.4 M_{\odot})^{0.47}}
\end{equation}
(electron degeneracy pressure and pair-production, which where neglected in this analysis, do not change this result substantially). The post-shock pressure is thus not dominated by radiation for
\begin{equation}\label{eq:rad_prs_dom}
    \rho_0 \gtrsim 6.6\times 10^5 \frac{ E_{51}^{1.4} (\mu/2)^{1.9}} {R^{1.6}_{8.5} (M/1.4 M_{\odot})^{0.89}}
    \DenUnit.
\end{equation}

Let us next estimate the evolution of the radiation to gas pressure ratio, $\eta$, with time. Assuming adiabatic evolution of an ideal gas in equilibrium with radiation, the density $\rho$ for which $\eta=1$ for a shell with post shock density $\rhosh=[(\gamma+1)/(\gamma-1)]\rho_0 $ and initial pressure ratio $\eta=\etsh$, is
\begin{equation}\label{eq:eta_cond}
    \rho/\rhosh = \etsh^{-1}e^{-8(\etsh -1)}.
\end{equation}
To estimate the time when the diffusion sphere reaches a layer with $\eta = 1$, we use the (time dependent) density and pressure profiles of the ejecta given in \citet{Rabinak10}, and use the method described there for determining the (time dependent) mass of the shell reached by the diffusion front. For time independent opacity we find
\begin{equation}\label{eq:t_drop_est_Rab10}
  \tdrop \approx 5.3 \times 10^3
  \frac{E^{0.83}_{51} R^{1.1}_{8.5} \kappa^{0.5}_{0.2} (\mu/2)^{1.44}}
    {f^{0.17}_{\rho} (M/1.4 M_{\odot})^{0.69}} \rm s,
\end{equation}
while for C/O opacity we find \citep[using the power-law approximation for the opacity suggested in][]{Rabinak10})
\begin{equation}\label{eq:t_drop_estCO_Rab10}
  \tdrop \approx 3 \times 10^3
    \frac{E^{0.66}_{51} R_{8.5} (\mu/2)^{1.2}}
    {f^{0.15}_{\rho} (M/1.4 M_{\odot})^{0.56}} \rm s
\end{equation}
(in both cases we neglected logarithmic corrections).

It is illustrative to compare $\tdrop$ obtained above to that obtained using the more detailed description of the ejecta profiles suggested by \citet{Piro10}. These profiles are obtained under the following assumptions: the pre-detonation envelope is in hydrostatic equilibrium and the pressure is dominated by degeneracy pressure (see eq.~\ref{eq:PresPrn15}), the velocity of the shock is given by eq.~(\ref{eq:vsAsRho0}), and the terminal velocity of a shell is $v_f = 2 v_s$ \citep[instead of $v_f = (6/7) v_s$ taken by][which is more appropriate for the planar phase]{Piro10}, the diffusion front is located at an optical depth $\tau = c/v_f$. The pressure in the ejecta is calculated using the relation $p = \psh \left(\rho/\rhosh \right)^{\gamma}$, where $\psh$ is the initial post shock pressure \citep[correcting Eq.~(17) of][which uses $\rho_0$ instead of $\rhosh$ in the denominator]{Piro10}.
For these profiles we find
\begin{equation}\label{eq:def_etaPiro}
     \etsh \equiv
     \frac{\prad}{\pgas}  = 910
     \frac{(\mu/ 2) v_9^{3/2} \rho_6^{0.27}}{(\rho_0/\DenUnit)^{0.52} },
\end{equation}
where $\vrun  = 10^9 v_9 $~cm sec$^{-1}$, and $\rrun = 10^6 \rho_6 \DenUnit$ are the velocity and density where the detonation wave transforms to a shock wave respectively. Thus, the assumption of radiation pressure dominance breaks for
\begin{equation}\label{eq:rad_prs_dom}
    \rho_0 \gtrsim 5\times 10^5 v_9^{2.9} \rho_6^{0.52}
    (\mu/2)^{1.9}\DenUnit.
\end{equation}
For constant opacity we find
\begin{equation}\label{eq:t_drop_est}
  \tdrop \approx 3.6 \times 10^3
  \frac{R_{8.5}^2 v_9^{2.16} \kappa_{0.2}^{0.5} \rho_6^{0.39} (\mu/2)^{1.8}}
      {\, (M/1.4 M_{\odot})^{0.5}} \rm s,
\end{equation}
while for C/O opacity we find
\begin{equation}\label{eq:t_drop_estCO}
  \tdrop \approx 2.2 \times 10^3
  \frac{R_{8.5}^{1.74} v_9^{1.76}  \rho_6^{0.31} (\mu/2)^{1.5}}
      {(M/1.4 M_{\odot})^{0.44}} \rm s.
\end{equation}

\subsection{Breakout flash}\label{sec:breakout_flash_ddt}

When the shock reaches layers with optical depth $\tau = \dm_{\rm BO} \kappa/(4\pi R_*^2)$ comparable to $c/v_s$, where $\dm_{\rm BO} $ is the mass exterior to the shell, photons outrun the shock and escape, producing a breakout flash. Our simulations' pre-detonation profiles (described in \sref{sec:dec_pre_det_prof}) are different from those used by \citet{Piro10}. However, as expected, the breakout energy is not sensitive to the details of the profiles. Using the simulation based (extrapolated) profiles derived in \sref{sec:extra_pre_det_prof}, and the shock description given in \sref{sec:sh_desc}, the shock velocity at $\tau =  c/v_s $ is mildly relativistic (for $\kappa_{0.2} = 1$). For mild-relativistic shocks, $v_s$ in Eq.~\eqref{eq:vsAsRho0} is replaced with $ \Gamma_s v_s$, where $\Gamma_s = [1-(v_s/c)^2]^{-1/2}$ \citep{TMM,Piro10}, and we find that the breakout energy is
\begin{equation}\label{eq:br_eng}
    E_{\rm BO} \approx 10^{40}
    \frac{R^{2.3}_{8.5} v_9^{1.16} f_\beta^{0.05} (2 \rho_6/\mu)^{0.21} }
        {(M/1.4 M_{\odot})^{0.16} (\Gamma_s/2.1)^{1.16} \kappa^{0.84}_{0.2} } \rm erg.
\end{equation}
$E_{\rm BO}$ depends weakly on $f_\beta\equiv\beta^{-4} (1-\beta)$, where $1-\beta = L/L_{edd}$, the ratio of the luminosity escaping the pre-detonation progenitor to the Eddington luminosity, is not accurately determined by our simulations. Our results is in agreement with previous estimates \citep{Imsh81,Piro10,Nakar10}, suggesting that the higher breakout energy obtained by \citet{Hoflich09} is due to the coarse numerical resolution near the stellar edge. The breakout flash is spread over $R_*/c$ (the photon diffusion time at shock breakout, which is smaller and is not weakly dependent on $ f_\beta $, has little effect on the observed properties of the emission). As shown by \citet{Budnik10,Katz10}, for mildly relativistic breakout velocity ($v/c>0.2$), a non-thermal spectrum extending to few hundred keV is expected.

\section{Numerical calculations}\label{sec:sim_desc}

We carried out one dimensional (1D) simulations of DDT SN Ia explosions, in order to obtain the outer ejecta profiles (i.e. density, pressure and temperature profiles), that determine the emission of radiation during the "spherical phase". Some aspects of burning, such as flame instabilities and turbulence, cannot be addressed in 1D simulations and are phenomenologically treated by using a parameterized deflagration velocity. Despite this caveat, we expect the simulations to describe the outer ejecta profiles truthfully.

The simulations have been performed using the Vulcan/1D (V1D) code \citep{Livne93}, that incorporates Lagrangian hydrodynamics, general EOS and nuclear burning. The EOS includes contributions from fully ionized gas and radiation at LTE. Free electron pressure and energy are computed using tabulated values of partially degenerate electron-positron gas. For burning we used the Alpha network of 13 elements from \elem{He}{4} to \elem{Ni}{56}. The nuclear reaction rates are taken from the NON-SMOKER database as described by \citet{Rauscher01}.

We assumed the "standard" scenario of thermonuclear runaway in a Carbon-Oxygen WD that approaches the Chandrasekhar mass by accretion from a companion. We assumed equal mass fractions for the two species, \elem{C}{12} and \elem{O}{16}. For a given central density and temperature we integrated the equation of hydrostatic equilibrium outwards under the assumption of constant entropy. We then modified the central density until the desired mass is achieved. Using variable zone masses we obtained a reasonable smooth mass distribution which goes down to $\sim 10^{-7} M_{\odot}$ in the outer cells. The total number of zones used was roughly 1000. The number of zones was increased to over 4000 in one simulation to test for convergence.

Our simulations start with a deflagration phase where the deflagration front propagates from the center of the progenitor outwards, with a parameterized local speed of the form used by \citet{Hoflich02}, where the deflagration velocity is take as the maximum between the Laminar deflagration velocity \citep[see Eq.~(43) of][]{Timmes92} and turbulent velocity.
For the turbulence velocity we use the parametrization suggested by \citet{Domnguez00}, $ f_{\rm turb} (g A L)^{1/2}$, where A is the Atwood number, $g$ is the gravitational acceleration, $L = (dr/dp) p $ is the characteristic pressure length scale and $f_{\rm turb} = 0.2$ and $0.4$ for "slow" and "fast" simulations respectively. The values for $f_{\rm turb}$ were calibrated from 3D calculations \citep{Khokhlov02}.

During the deflagration phase, the progenitor expands and its outer layers move with a typical velocity of 3 to $6 \times 10^8 \VelUnit$ depending on the deflagration velocity. In regions where the expansion velocity is higher than the matter sound speed, weak shock waves form and heat the plasma. This leads to modifications of the density and pressure profiles of the progenitor, as discussed in detail in \sref{sec:dec_pre_det_prof}. The fully ionized gas EOS is not valid at the outer parts of the progenitor, leading to inaccurate initial profiles in these regions. These profiles, however, are modified by the heating during the deflagration phase, and therefore have little effect on the final ejecta profiles.

The deflagration front propagates until it reaches a critical density $\rocrit$, where we induce a detonation wave by increasing the deflagration speed to nearly sonic over few zones. Three values are taken for $\rocrit$: 1, 2 and 2.5$ \times 10^7 \DenUnit$. These are denoted by 1, 2 and 2.5 in the simulation names. The detonation front propagates outwards until the entropy produced by the burning of elements becomes smaller than the entropy that is produced by the compression, at which point the detonation transforms into a shock wave.
The transition of the detonation to a shock wave as well as the shock propagation are discussed in detail in \sref{sec:sh_desc}.
{We follow the expansion of the progenitor after the shock wave reaches the stellar edge, neglecting the effects of radiation transport.

\begin{table}[ht]
\caption{Simulation Parameters}\label{table:sim_param}
\centering 
\begin{tabular}{lcccccc} 
\hline\hline 
Name${}^a$  &Color    &$\tdet^b$ &$K_{ch}{}^c$    &$\gamma_p{}^c$ &$R_*^d$    &$E_{51}^{e}$ \\ [0.5ex] 
&               &[s]       &                &               &[$10^8$cm]  \\ [1ex]
\hline 
DDT1s   &blue   &3.05       &3.5            &0.95           &5.3        &1.27\\ [1ex]
DDT2s   &green  &2.8        &7.1            &0.89           &4.3        &1.52\\ [1ex]
DDT2.5s &red    &2.75       &7.9            &0.89           &4.1        &1.55\\ [1ex]
DDT1f   &cyan   &1.75       &1.3            &1              &4.9        &1.31\\ [1ex]
DDT2f   &magenta &1.6       &6.7            &0.86           &3.9        &1.55\\ [1ex]
DDT2.5f &black  &1.6        &6.7            &0.86           &3.9        &1.58\\ [1ex]
\hline 
\multicolumn{7}{l}{${}^a$ 1, 2 and 2.5 correspond to DDT transition at $\rocrit =$} \\
\multicolumn{7}{l}{${}^{\phantom a}$ $ 10^7 \DenUnit \times $ 1, 2 and 2.5. "s" and "f" stand for slow and}\\
\multicolumn{7}{l}{${}^{\phantom a}$ fast deflagration velocities (cf.~\sref{sec:sim_desc})}\\
\multicolumn{7}{l}{${}^b$ The time at which detonation is induced.} \\
\multicolumn{7}{l}{${}^c$ Fit parameters for $P_0(\rho_0)$ at $t_s = \tdet $ and } \\
\multicolumn{7}{l}{${}^{\phantom c}$  $\rho_0 < 10^4 \DenUnit$ (eq.~\eqref{eq:PresPrSim}).}\\
\multicolumn{7}{l}{${}^d$ Progenitor radius at $t = \tdet $.} \\
\multicolumn{7}{l}{${}^{e}$ Total ejecta energy in units of $10^{51}$~erg.} \\
\hline 
\end{tabular}
\end{table}

\section{UV/optical emission during the spherical phase}\label{sec:res_early_n_BO}

\subsection{Density and pressure profiles of the outer parts of the ejecta}\label{sec:fin_prof}

We derive the emission of radiation using the density and temperature profiles at the end of the simulations, $t_s = 100$~s (note that $t_s$ is measured from the onset of deflagration, while we use $t$ to denote the time measured from breakout). At this stage the ejecta has expanded significantly and the radius $r$ of a relevant mass shell is $\gg R_*$ and is approximately given by $ t v_f $, where the shell's velocity $v_f $ is approximately its terminal velocity. The simulations' mass resolution at the outer cells, $\sim 3\times 10^{-7} M_{\odot}$, sets a lower limit to the time from which the emission can be reliably calculated based on the simulations, $t\gtrsim 3\times 10^2$s.

We neglect the reheating by the radioactive decay of elements synthesized in the explosion process (valid for the outer layers of the ejecta, in which only a small fraction of the  elements are fused), and extrapolate the profiles to later times assuming that the ejecta expand adiabatically and that the fluid elements reached their final velocity. We use an EOS of radiation in equilibrium with an ideal gas ($\mu = 1.7455$ for fully ionized equal mass fractions of C and O). Note that although we assume full ionization for the EOS, we do not make this assumption for the opacity. This approximation for the EOS is accurate as long as the thermal energy density is dominated by radiation, and becomes less accurate when the plasma pressure becomes significant. This implies that our light curves are not accurate at $t\gtrsim \tdrop$.

\begin{figure}
\includegraphics[scale=1]{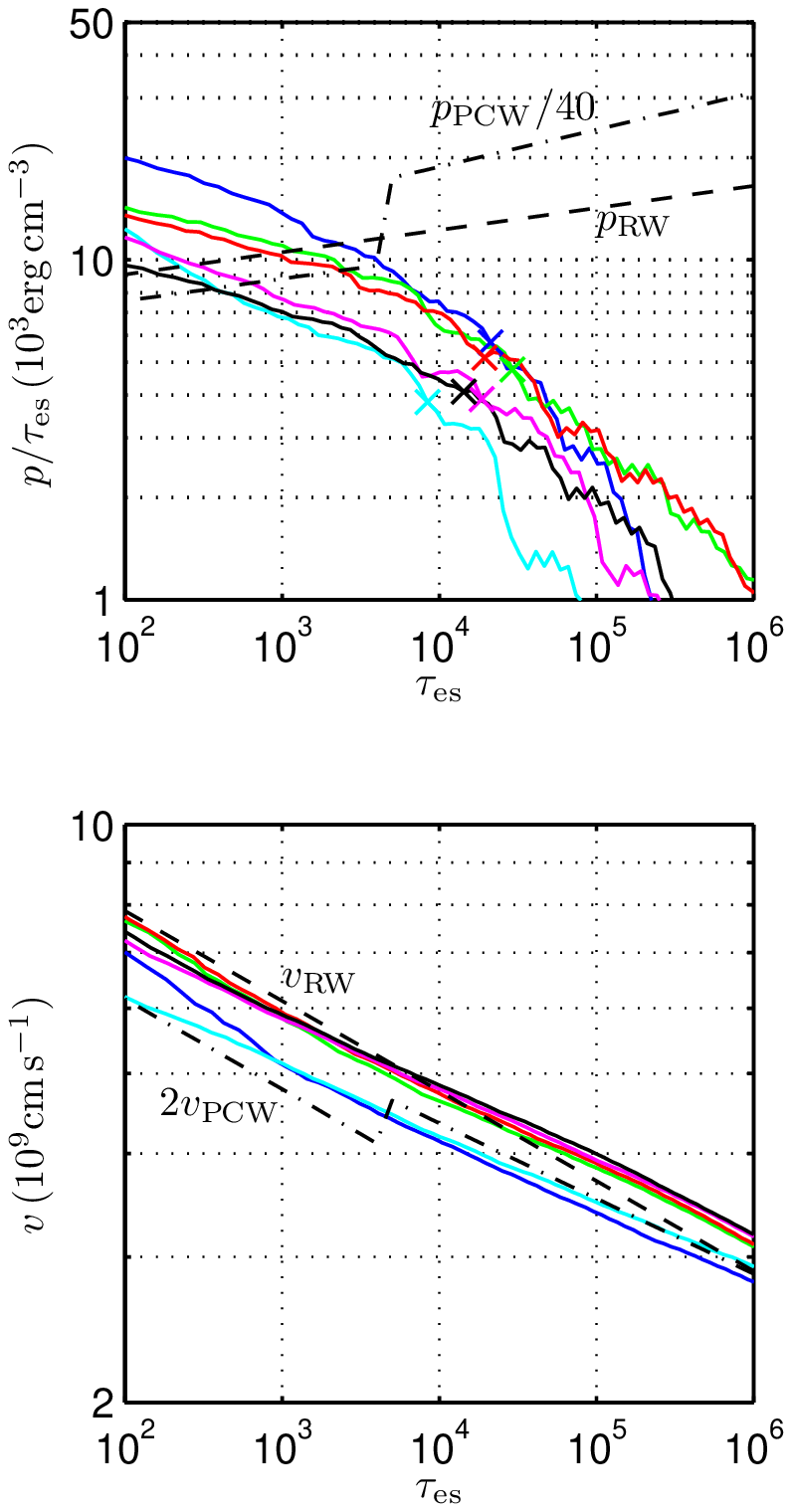}
\caption{\label{fig:FinPrsFitAsTau}
Pressure and velocity as function of electron-scattering optical depth obtained from the simulations (solid lines) at $t_s=100$s, compared to the analytic estimates of \citet{Rabinak10} (dashed lines) and \citet{Piro10} (dash-dotted, corrected as explained in the text preceding Eq.~\eqref{eq:def_etaPiro}). Line colors correspond to different simulation parameters as given in Table \ref{table:sim_param}. "x" marks indicate the position where the radiation and gas pressure are equal.
}
\end{figure}

In Fig.~\ref{fig:FinPrsFitAsTau} we compare the pressure and velocity profiles at the end of the simulations to the profiles suggested by \citet{Rabinak10} (for $n=3$) and \citet[][corrected as explained in the text preceding Eq.~\eqref{eq:def_etaPiro}]{Piro10}, denoted by RW and PCW subscripts respectively. For the model of \citet{Rabinak10} we used $E_{51} =1,  R_{8.5}=1, {\rm and \;}  M = 1.4 M_{\odot}$, similar to the values obtained from the simulations (see table \ref{table:sim_param}). The discontinuity in the PCW profiles results from a transition between regions with different pre-detonation density profiles. Both the RW and PCW models do not properly describe the pressure behavior at large optical depth, where the pressure is no longer purely dominated by radiation and thus drops faster with time, as explained in \S~\ref{sec:est_t_drop}.

For the lower deflagration velocity the pressure is higher by $\sim 70\%$ at $10^2 \lesssim \tes \lesssim 10^4$, where $\tes$ is the optical depth for electron scattering ($\kappa_{0.2} = 1$). This difference is due to the fact that for the low velocity explosions the envelope is heated to higher temperatures during the deflagration phase. Thus, for a given pre-detonation density $\rho_0$ the mass fraction $\dm$ (integrated mass to the stellar surface) is larger compared to that in high velocity explosions. For $\gamma = 4/3$, as $\tes \propto \dm (t v_s)^{-2} $ and the density in the ejecta $\rho \propto \dm (t v_s)^{-3} $, the pressure in the ejecta $p \propto v_s^2 (\rho/\rho_0)^{\gamma} \propto (\dm/\rho_0)^{1/3} \tes t^{-2}$, implying that for lower $\rho_0$ at given $\dm$ the pressure is larger at a given $\tes$.

\subsection{$L$ and $T_{\rm eff.}$}\label{sec:early emission_ddt}

We use the density and pressure profiles obtained as described in \S~\ref{sec:fin_prof} to derive the properties of the emitted radiation. Due to the deviation from pure radiation pressure domination, the pressure profiles deviate significantly from $p\propto\tau$ (which is valid for the core collapse progenitors). This implies that photon diffusion may have a non negligible effect on the predicted luminosity \citep{Rabinak10}. We therefore estimate the luminosity at time $t$ as $dE/dt$, where $E(t)$ is the radiation energy that escapes by diffusion up to time $t$.
The diffusion depth is determined as the depth at which $\tau = c/v_f$, and $\tau $ is calculated with the opacity taken from the OP project tables \citep{OPCD}. The effective temperature is calculated as $T_{\rm eff}=(L/4\pi \rph^2 \sigma)^{1/4}$, where $\rph$ is the radius of the photosphere ($\tau = 1$). The temperature at the photosphere ($\tau = 1$) is typically larger than $T_{\rm eff}$ by $10\%\div 20\%$, which implies that neglecting diffusion leads to an overestimate of $L$ by a factor of $1.5\div2$. The simulations' resolution was not fine enough to allow an accurate determination of the color temperature.

\begin{figure}
\includegraphics[scale=1]{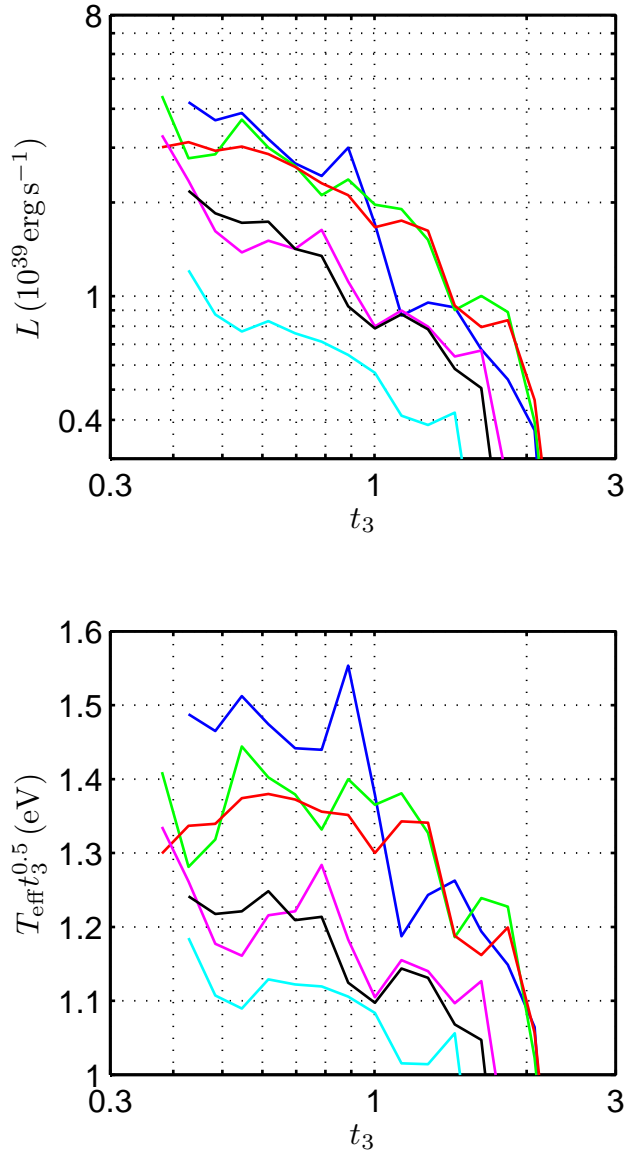}
\caption{\label{fig:LcSim2TheoComp}
Bolometric luminosity and effective temperature derived from the simulations. Line colors correspond to different simulation parameters as given in Table \ref{table:sim_param}. $t = t_3 10^3$s.
}
\end{figure}

The resulting luminosities and effective temperatures are given in Fig.~\ref{fig:LcSim2TheoComp}. The luminosity depends weakly on $\rocrit$, and is lower for larger deflagration velocities, reflecting the lower pressure at given $\tes$ (see Fig.~\ref{fig:FinPrsFitAsTau}). Comparing the results presented in the figure with eqs.~(\ref{eq:L_n3CO}),~(\ref{eq:T_n3CO}) and~(\ref{eq:t_drop_estCO_Rab10}), we find that the simple model of \S~\ref{sec:spherical} provides a good description of the properties of the emitted radiation. For the slow deflagration velocity the luminosity is similar to that predicted by the simple analytic analysis, while for the larger velocity it is 2-3 times lower.

An estimate of the luminosity expected for a pure He composition of the outer shells may be obtained by extrapolating the simulation profiles assuming an EOS with $\mu = 2$, and calculating the emission using the opacity of He. The luminosity obtained in this case is $\sim 50\%$ lower and the effective temperature is $\sim 15\%$ lower than the results obtained for C/O, consistent with the simple model of \S~\ref{sec:spherical}.

\section{summary and discussion}\label{sec:sum_and_disc}

We have presented in \S~\ref{sec:spherical} a simple model of the UV/optical emission during the spherical phase, assuming a power-law density profile of the progenitor that is not modified during the deflagration phase, and a self-similar shock propagation, ignoring the fact that part of the explosion energy is released by nuclear burning at large radii. The predicted luminosity and effective temperature are given in eqs.~(\ref{eq:L_n3})-(\ref{eq:T_n3CO}) for different compositions of the outer shells of the progenitor. For $t>300$~s the photosphere propagates beyond the outer $\sim 10^{-5} M_{\odot}$, the estimated maximum mass of a possible H shell \citep{Shen09a},  and the effective temperature drops below 3~eV. At this time we therefore expect the radiation to be emitted from shells dominated by He or C/O, with luminosity and temperature given by eqs.~\eqref{eq:L_n3He} and~\eqref{eq:T_n3He} or \eqref{eq:L_n3CO} and \eqref{eq:T_n3CO} respectively. At earlier times the detailed properties of the emission depend on the mass fraction of H. The simple model predicts a nearly time independent luminosity, $L\simeq10^{39.5}{\rm erg/s}$, at $300\,{\rm s}<t<10^3$~s, and a strong suppression of the flux at $t>\tdrop\sim1$~hr (see eqs.~\ref{eq:t_drop_est} and~\ref{eq:t_drop_estCO}) due to the deviation from pure radiation domination.

Comparing the results of the model to those of detailed 1D simulations, we find that it provides an acceptable description of the velocity profile of the ejecta, and a rough description of the pressure profile in regions where radiation dominates the pressure (see fig.~\ref{fig:FinPrsFitAsTau}; a detailed discussion of the pre-detonation profile, which is significantly affected at $ \rho_0 \lesssim 3\times 10^4 \DenUnit$ by weak shocks during the deflagration phase, is given in the appendix). Thus, we expect the model to provide an acceptable description of the luminosity at $t<\tdrop$.

Indeed, we find that eqs.~(\ref{eq:L_n3CO}),~(\ref{eq:T_n3CO}) and~(\ref{eq:t_drop_estCO_Rab10}) provide an acceptable description of the properties of the emitted radiation. We find that the luminosity depends weakly on $\rocrit$, the density at which detonation is initiated, and is lower for larger deflagration velocities. For the slow (turbulent) deflagration velocity (see \S~\ref{sec:sim_desc} for a description of the turbulent velocity prescription) the luminosity is similar to that predicted by the simple analytic analysis, while for the larger velocity it is 2-3 times lower (see fig.~\ref{fig:LcSim2TheoComp}).

As expected, we find that the breakout emission during the planar phase is not sensitive to the density profile modifications. We find $E_{\rm BO} \approx 10^{40}$~erg (see eq.~\ref{eq:br_eng}), spread over $R_*/c$, in agreement with previous estimates \citep{Imsh81,Piro10,Nakar10}. As shown by \citet{Budnik10,Katz10}, for the predicted mildly relativistic breakout velocity ($v/c>0.2$), a non-thermal spectrum extending to few hundred keV is expected.

\acknowledgements We thank A. Glasner and D. Kushnir for useful discussions. This research was supported in part by ISF, UPBC, GIF and Minerva grants.\\

\appendix

\section{Analysis of numerical simulations' results}\label{sec:model4fin_prof}

\subsection{Pre-detonation pressure and density profiles}\label{sec:dec_pre_det_prof}

We consider the pressure and density profiles of the outer layers of the progenitor, which are modified by weak shocks generated during the deflagration phase, focusing on the profiles at $t = \tdet$. During the deflagration phase the progenitor WD expands. In regions, where the expansion velocity is similar to or higher than the sound speed, weak shock waves form, which heat the envelope and modify its pressure profile. During this phase, two physical processes affect the pressure profile: shock wave heating and diffusion of energy. Diffusion affects significantly regions in which the photon diffusion time (to the stellar surface) is similar to the deflagration time. The outer part of the progenitor's envelope may be divided into three regions distinguished by the physical processes which affect them: (1) The inner region, which is negligibly affected by shock heating and diffusion; (2) The intermediate region, which is affected by shock heating but is negligibly affected by diffusion; (3) the outer region, which is affected by both shock heating and photon diffusion. The mass resolution in our simulations does not allow us to fully resolve region (2). We denote by (2A) and (2B) the parts of this region which are resolved and unresolved by the simulations respectively. Below we describe the simulations' results for regions regions (1) and (2A). The unresolved regions, (2B) and (3), are discussed in \sref{sec:extra_pre_det_prof}.

Let us first discuss the pressure profile expected at the outer shells of the progenitor based on simple theoretical arguments. For simplicity, we neglect the shells' self-gravity and thickness (relative to the progenitor radius $R_*$ at the onset of detonation). We denote the density of a shell at the onset of detonation by $\rho_0$. Assuming that the star is in hydrostatic equilibrium, we have for the outer layers
\begin{equation}\label{eq:DmP0relation}
  \dm(\rho_0) \approx \frac{4\pi R_*^4 P_0(\rho_0)}{G M},
\end{equation}
where $M$ is the total mass of the progenitor, $P_0$ is the pressure, and $\dm$ is the integrated mass from the point where the density equals $\rho_0$ out to the stellar surface. For a polytropic EOS, $P_0 \propto \rho_0^{1+1/n}$, the density profile is $\propto \delta^{n}$, where $\delta = (R_*/r -1)$.

We consider first the inner region (1). In this region the pressure is negligibly affected by the deflagration process and is therefore dominated by the electron degeneracy pressure. In this case $P_0(\rho_0)$ is polytropic with index $n = 3/2 $, and may be approximated by \citep[following][]{Piro10}
\begin{equation}\label{eq:PresPrn15}
    P_{\rm deg} \equiv 9.91 \times 10^{12} \left(\frac{\rho_0}{\mu_e\DenUnit}\right)^{5/3} \PresUnit ,
\end{equation}
where $\mu_e$ is the molecular weight per electron. The sound speed in these layers is $c_s = 4.1 \times 10^{6} (\rho_0/\DenUnit)^{1/3} \mu_e^{5/6} \VelUnit$.

Let us consider next region (2A). The simulations show that when the deflagration front reaches a density $\rocrit \sim 10^7 \DenUnit$, at time $\tdet$, the outer layers of the progenitor expand at a speed $3\div6 \times 10^8 \VelUnit$. The pressure profile deviates from that of region (1) at densities $\sim 3\times 10^4 \DenUnit$, beyond which it becomes flatter and nearly linear in $\rho_0$. This change in the density profile is the result of shock heating during expansion. We use the simulations to obtain an approximation for $P_0(\rho_0)$. In Table \ref{table:sim_param} we give the best fit parameters for a pressure profile of the form
\begin{equation}\label{eq:PresPrSim}
   P_0(\rho_0)= K_{ch}\times 10^{15}  \left(\frac{\rho_0}{\DenUnit}\right)^{\gamma_p} \PresUnit,
\end{equation}
for $t = \tdet $ and $ \rho_0 < 10^4 \DenUnit$. In all of the fits the index $\gamma_p$ is in the range $0.9\div 1$, implying a nearly isothermal profile, in agreement with the simulations of \citet[][see figure 1 in that manuscript]{Hoflich09}, which have a similar resolution to that of our simulations, $<10^{-6} M_{\odot}$.

The pressure in region (2A) is different than that assumed by \citet{Imsh81,Piro10,Nakar10} (for the same $\rho_0$ range). In particular, it corresponds to a negative polytropic index ($n<0$) and does not correspond to a density profile declining as a power of $\delta$ (distance from the progenitor's edge). The density profile we find falls approximately exponentially with $ r $, similar to the profile produced by an isothermal gas. The shells of region (2A) determine the emission on time scales of minutes to hours.

In the left panel of Fig.~\ref{fig:dmAsRho0FP} we compare the results of the simulations to the approximations of Eqs.~\eqref{eq:PresPrn15} and \eqref{eq:PresPrSim}. Since the profile is nearly isothermal at the outer regions, we normalize $P_0$ to
\begin{equation}\label{eq:PresThrm}
  P_{\rm thrm}(\rho_0) \equiv 4.79\times 10^{15} \left(\frac{\rho_0}{\DenUnit}\right) \frac{ \Ttkev }{\mu/2}\, \PresUnit,
\end{equation}
where $T = 10 \Ttkev$~keV and $\mu$ is the molecular weight.
The approximation of Eq.~\eqref{eq:PresPrn15} is good to a few percent in the density range $3\times 10^4 \DenUnit< \rho_0< 10^6 \DenUnit$, and the approximation of Eq.~\eqref{eq:PresPrSim} is good to tens of percent for $\rho_0 < 10^4 \DenUnit$. In the latter density range, $P_0$ is given by $P_{\rm thrm}$ to within a factor of 2.

\begin{figure}
\includegraphics[scale=1]{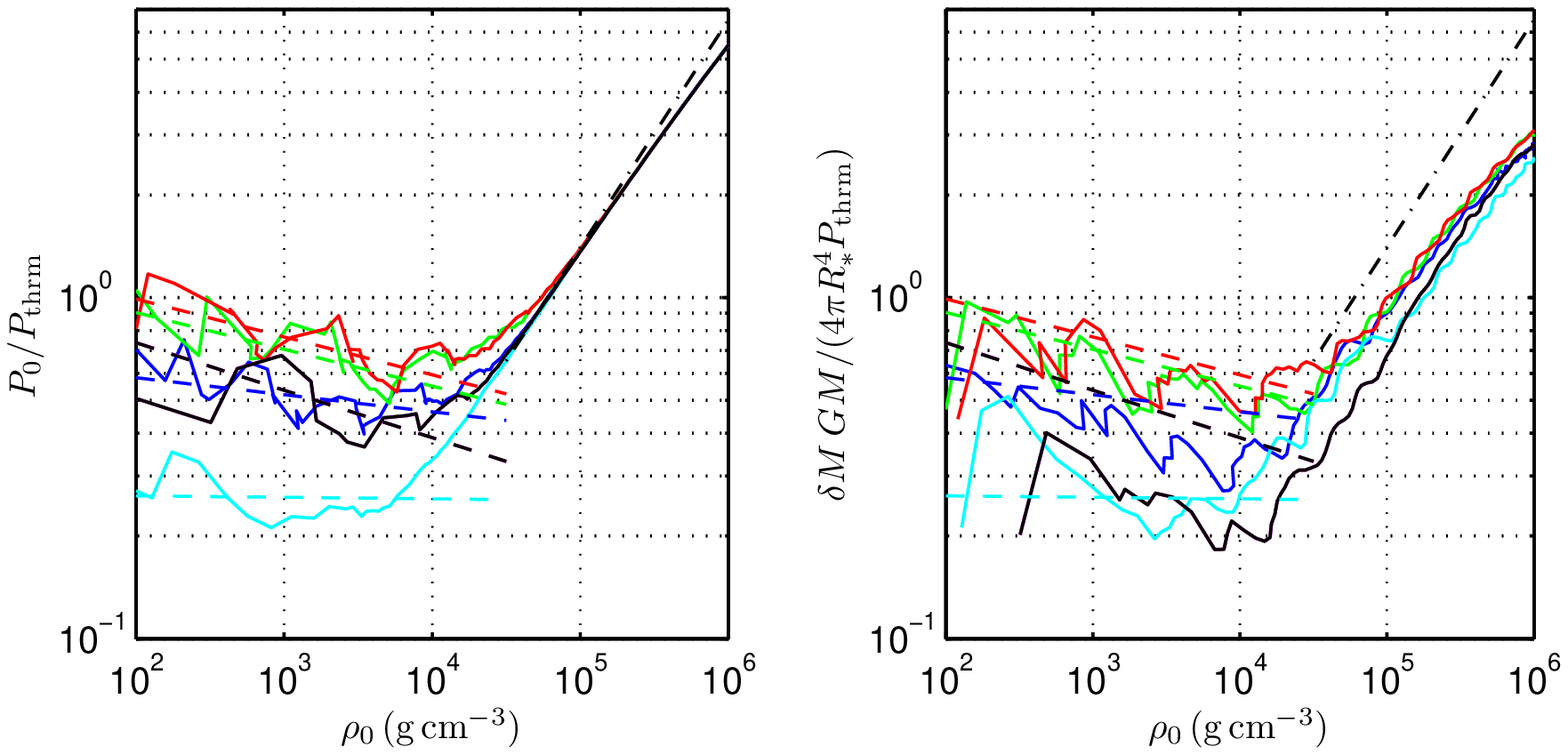}\label{fig:dmAsRho0FP}
\caption{
Left: $P_0(\rho_0)$ obtained in the simulations at $t = \tdet $ (solid lines) compared to the approximations of Eq.~\eqref{eq:PresPrn15} (black dash dotted line), and Eq.~\eqref{eq:PresPrSim} (with the values given in Table \ref{table:sim_param}, dashed lines). The pressure is normalized to the pressure given by Eq.~\eqref{eq:PresThrm} for $\Ttkev = 1$ and  $\mu = 2$.
Right: $\dm(\rho_0) $ obtained in the simulations at $t = \tdet $ (solid lines) compared to the approximation of Eq.~\eqref{eq:DmP0relation}, with $P_0$ given by Eq.~\eqref{eq:PresPrn15} (black dash dotted line) or Eq.~\eqref{eq:PresPrSim} (dashed lines).
The masses are normalized to the mass given by Eq.~\eqref{eq:DmP0relation} with $P_0 = P_{\rm thrm}$ (with $\Ttkev = 1$ and  $\mu = 2$).
Different colors correspond to different simulation parameters as specified in Table \ref{table:sim_param}.
}
\end{figure}

In the right panel of Fig.~\ref{fig:dmAsRho0FP} we compare $\dm(\rho_0)$ obtained in the simulations to that given by Eq.~\eqref{eq:DmP0relation} using the pressure approximations of Eqs.~\eqref{eq:PresPrn15} and \eqref{eq:PresPrSim} (with the values given in Table \ref{table:sim_param}). Eq.~\eqref{eq:DmP0relation} with $P_0$ given by Eq.~\eqref{eq:PresPrn15} is accurate to a factor $\sim 2$ in the density range $3\times 10^4 \DenUnit <  \rho_0 < 10^6 \DenUnit$. The discrepancy is partially due to deviations from hydrostatic equilibrium and partially due to the fact that $R_*$ over estimates $r$ by up to 15\% in this density range. Eq.~\eqref{eq:DmP0relation}, with $P_0$ given by Eq.~\eqref{eq:PresPrSim}, is accurate to a few tens of a percent in the density range $\rho_0 < 10^4 \DenUnit$. In this density range, substituting $P_{\rm thrm}$ for $P_0$ gives an approximation for $\dm $, which is good to up to a factor of 2.

\subsection{Shock propagation}\label{sec:sh_desc}

As the detonation wave propagates in the declining density profile of the star, the entropy produced by the burning of elements decline, until eventually this entropy is smaller than the entropy that is produced by the shock compression. At that point the detonation wave transforms to a shock wave. The shock wave continues to accelerate in the declining density profile. In regions where the density falls as $\delta^{n},$ shock acceleration is described by the Gandel'Man-Frank-Kamenetskii--Sakurai self similar solutions \citep{GandelMan56,Sakurai60}. In regions where the density falls exponentially with $\delta$, the acceleration is described by the self similar solution of \citet{Grover66}. In both types of solutions, the shock velocity is given by
\begin{equation}\label{eq:vsAsRho0}
    v_s = \vrun \left(\frac{\rho_0}{\rrun}\right)^{-\beta_1}.
\end{equation}
For a radiation dominated shock, with post shock adiabatic index of $\gamma_s = 4/3$, $\beta_1 \approx 0.19 $ for density profiles declining as a power law with $n = 3, 3/2$ \citep{Grassberg81,MM}, and $\beta_1 \approx 0.176 $ for exponential density profiles \citep{Grover66}. The self-similar solutions also show that the terminal velocity of the shells, $v_f$, is $\propto v_s $. For power-law density profiles $v_f \cong 2 v_s$, while for exponential density profiles $v_f \cong 1.5 v_s$.

The transition from a detonation wave to a shock wave was analyzed by \citet{Piro10}. The transition density $\rrun $ was defined as the density at which the "induction length" $\lambda = (q/\epsilon)\vdet,$ where $q, \epsilon$ are the energy and energy generation rate of the detonation process respectively, is smaller than the scale over which the progenitor density changes, assuming  $\vdet$is given by the Chapman-Jouguet detonation velocity at the corresponding density. For a WD equally composed of $^{12}$C and $^{16}$O they found $\rrun \approx 2\times 10^6 \DenUnit$, in agreement with our simulations (although the equation for $\epsilon$ was take in \citet{Piro10} at the wrong electron screening regime.) It should be noted here that the values of $\vrun, \rrun $ may be affected by instabilities which are not considered  here \citep[e.g.~][]{Dominguez11}.

In the left panel of Fig.~\ref{fig:VsRho0} we compare the shock velocity and the terminal velocity obtained in a simulation to the approximation of Eq.~\eqref{eq:vsAsRho0}. This figure shows that although the density profile changes from a power of $\delta$ to an exponential, the approximation Eq.~\eqref{eq:vsAsRho0} holds. The figure also shows that the values estimated analytically for $\rrun$ and $\vrun$ are accurate to $\sim 10\%$. The ratio $v_f/v_s$ is not independent of $\rho_0$, and declines from a ratio of $\approx 2.4$ at $\rho_0 = 10^5 \DenUnit$ to $\approx 1.4$ at $\rho_0 = 10 \DenUnit$. The decline in the ratio $v_f/v_s$ with $\rho_0$ can be attributed to the transition from a power-law to an exponential density profile and to spherical affects \citep[see~][]{MM}. In the right panel of Fig.~\ref{fig:VsRho0} we compare the velocity at $t_s = 100$~s (terminal velocity) to the approximation
\begin{equation}\label{eq:vfAsRho0}
  v_f = 2.5 f_v \, v_s (\rho_0/\rrun)^{0.05}.
\end{equation}

\begin{figure}
\includegraphics{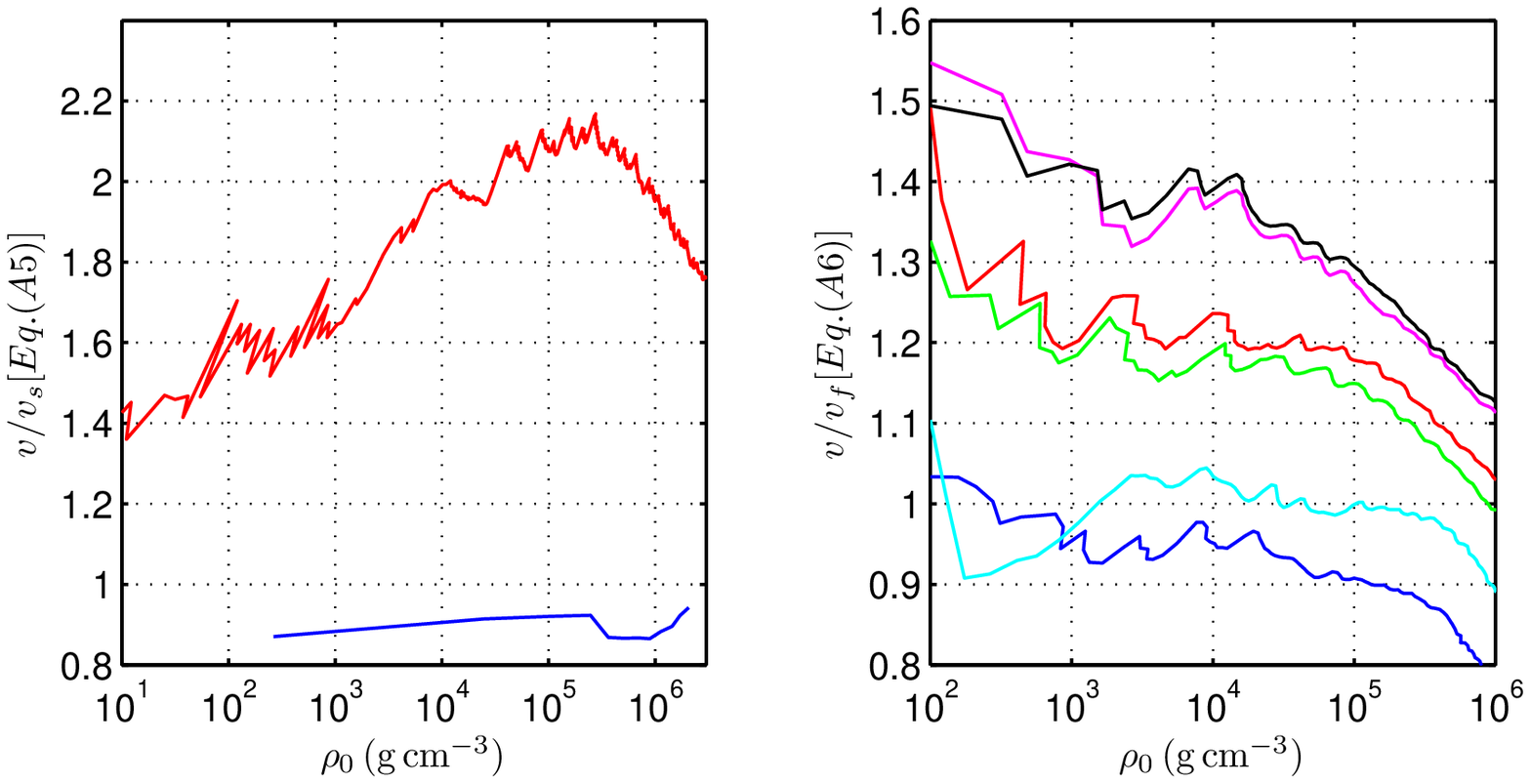}
\caption{\label{fig:VsRho0}
Left: Shock velocity (blue) and terminal velocity ($t_s = 100$~s, red) obtained in one simulation (with higher dumping resolution),
normalized by the shock velocity given by Eq.~\eqref{eq:vsAsRho0}, with $\beta_1 = 0.18 $.
Right: Terminal velocities ($t_s = 100$~s, red) obtained in the simulations, normalized to the approximation of Eq.~\eqref{eq:vfAsRho0} with $\beta_1 = 0.18 $ and $f_v = 1$. Different colors correspond to different simulation parameters as specified in Table \ref{table:sim_param}.
}
\end{figure}

\section{Extrapolating the pre-detonation profiles to $\dm_{\rm BO}$}\label{sec:extra_pre_det_prof}

Our simulations do not resolve regions with densities $\rho_0 \lesssim 10^2 \DenUnit $ at $\tdet$ and we only consider them qualitatively.
These regions determine the emission at $t\lesssim 3\times 10^2$s (cf.~\sref{sec:early_emis_sh_heated}) and have little effect on the emission at later times. Assuming that the pressure profile given by Eq.~\eqref{eq:PresPrSim} continues to densities $\ll 10^2 \DenUnit$, the pressure becomes dominated by radiation at $\rho_0  \sim 1 \DenUnit$, the flux reaches $\sim g c/ \kappa $ and the luminosity approaches the Eddington luminosity, $L_{edd}.$ In this case, the luminosity is $ L \approx L_{edd} = 4\pi c G M/ \kappa \sim 10^{38} \LumUnit$, which is similar to the luminosity obtained by \citet{Hoflich09} before the transition to detonation, and is larger by 3 orders of magnitude than the luminosity assumed by \citet{Piro10}.
The simulations indicate therefore that the heating process is efficient in the sense that it leads to $L\simeq L_{edd}$ during the deflagration phase. that this upper bound is reached.

It is reasonable to assumed that near the stellar surface, in regions where the diffusion time to the stellar surface is $ \ll \tdet$, the luminosity is constant. Assuming that the envelope is at hydrostatic equilibrium, that opacity is constant and that the pressure is a sum of a radiation pressure and an ideal gas pressure with mean molecular weight $\mu$, we have in this region
\begin{equation} \label{eq:PresPrn3}
    P_0 =
    10^{15} \left(\frac{1-\beta}{\beta^4 }\right)^{1/3} \left(\frac{2 \rho_0}{\mu \DenUnit}\right)^{4/3} \PresUnit,
\end{equation}
where $1-\beta = L/L_{edd}$. The large $L$ obtained in the simulations indicates that  $f_\beta\equiv \beta^{-4} (1-\beta) \gtrsim 1$.
This pressure profile is similar to the one suggested by \citet{Piro10} for the outermost layers but its normalization is higher by an order of magnitude. For this profile, the diffusion time to stellar edge is
\begin{equation}
    \tdiff \approx 2 \frac{f_\beta^{2/3} R_{8.5}^4
    \kappa_{0.2}}{ (M/1.4 M_{\odot})^2 (\mu/2)^{8/3}}
    \left(\frac{\rho_0}{\DenUnit}\right)^{5/3} \sec,
\end{equation}
where $\kappa = 0.2 \kappa_{0.2} \OpUnit$ is the opacity.
The density for which $\tdiff \approx \tdet $ is roughly $\rho_0 \sim 1 \DenUnit$ (for $\tdet$  given in Table \ref{table:sim_param}).

\bibliographystyle{apj}
\bibliography{general}

\end{document}